\documentclass{webofc}
\usepackage[varg]{txfonts}   % Web of Conferences font
\begin{document}
\title{Potentiality of automatic parameter tuning suite available in ACTS track reconstruction software framework}

\author{\firstname{Rocky Bala} \lastname{Garg}\inst{1}\fnsep\thanks{\email{rocky.bala.garg@cern.ch}} \and
        \firstname{Corentin} \lastname{Allaire}\inst{2}\fnsep\thanks{\email{corentin.allaire@cern.ch}} \and
        \firstname{Andreas} \lastname{Salzburger}\inst{3} \and
        \firstname{Hadrien} \lastname{Grasland}\inst{2} \and
        \firstname{Lauren} \lastname{Tompkins}\inst{1} \and
        \firstname{Elyssa} \lastname{Hofgard}\inst{1}
}

\institute{Stanford University 
\and
           Universit\'e Paris-Saclay
\and
           CERN
          }

\abstract{%
  Particle tracking is among the most sophisticated and complex part of the full event reconstruction chain. A number of reconstruction algorithms work in a sequence to build these trajectories from detector hits. Each of these algorithms use many configuration parameters that need to be fine-tuned to properly account for the detector/experimental setup, the available CPU budget and the desired physics performance. Few examples of such parameters include the cut values limiting the search space of the algorithm, the approximations accounting for complex phenomena or the parameters controlling algorithm performance. The most popular method to tune these parameters is hand-tuning using brute-force techniques. These techniques can be inefficient and raise issues for the long-term maintainability of such algorithms. The open-source track reconstruction software framework known as “A Common Tracking Framework (ACTS)” offers an alternative solution to these parameter tuning techniques through the use of automatic parameter optimization algorithms. ACTS come equipped with an auto-tuning suite that provides necessary setup for performing optimization of input parameters belonging to track reconstruction algorithms. The user can choose the tunable parameters in a flexible way and define a cost/benefit function for optimizing the full reconstruction chain. The fast execution speed of ACTS allows the user to run several iterations of optimization within a reasonable time bracket. The performance of these optimizers has been demonstrated on different track reconstruction algorithms such as trajectory seed reconstruction and selection, particle vertex reconstruction and generation of simplified material map, and on different detector geometries such as Generic Detector and Open Data Detector (ODD). We aim to bring this approach to all aspects of trajectory reconstruction by having a more flexible integration of tunable parameters within ACTS.
}
\maketitle
\section{Introduction}
 The reconstruction of trajectories for charged particles is a computationally demanding yet vital aspect of any high-energy physics experiment. The computational overload multiplies quadratically with the number of particles in the detector. Future progress demands dedicated efforts to enhance algorithmic performance while managing computational overhead. A viable avenue for such an enhancement involves optimizing algorithmic parameters that span from particle properties preselection to simplifying detector effects and limiting search spaces. However, the sheer number of such optimizable parameters, and their dependence on the specific detector configurations and experimental circumstances requires a significant amount of work and regular updates.
 
 This paper puts forth the proposal to utilize the auto-tuning methodologies for the optimization of parameters within different tracking algorithms. The effectiveness of these techniques has been demonstrated through the development of the simplified material models for detectors (known as material mapping) and the optimization of parameters in track seeding and primary vertex reconstruction algorithms. These methodologies offer the potential to identify an optimal parameter set based on the prevailing experimental conditions, all while minimizing human intervention.

\section{ACTS and dataset}
\begin{figure}[!h]
\begin{center}
    \includegraphics[width=0.6\textwidth]{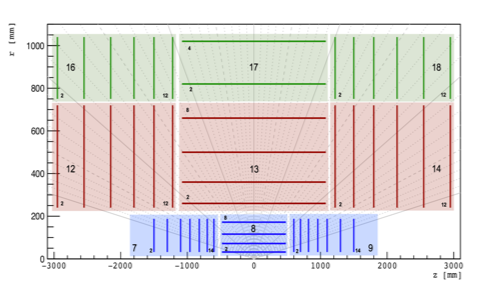}
    %$\includegraphics[width=0.45\textwidth]{Plots/ODD.png}
    \caption{Layout of ACTS Generic Detector. The Open Data Detector share similar layout but have more realistic interaction material.}
    \label{fig:ACTS_det}
    \end{center}
\end{figure}

	A Common Tracking Software (ACTS)~\cite{Ai:2021ghi} is a tracking framework being developed since 2016 as an international collaboration with the goal of providing a generic, experiment-independent open-source software framework for charged particle tracks reconstruction. To facilitate comprehensive research, ACTS offers two simulated detector geometries: the Generic Detector 
 as shown in figure~\ref{fig:ACTS_det} is a standard all-silicon LHC-type tracking detector used in the TrackML~\cite{Amrouche2023} challenge and the Open Data Detector (ODD)~\cite{corentin_allaire_2022_6445359}, an evolved version of Generic Detector implemented using DD4Hep~\cite{Frank_2014} that provides the support structure and the material akin to a real detector. The Generic detector has been used for the seeding and vertexing optimization studies while ODD has been used for the material mapping studies in this paper.
 The adaptability of ACTS to various detector geometries positions it as an ideal platform for researching and testing auto-tuning techniques with different algorithmic configurations. %The optimization algorithms featured in this paper have been integrated into the ACTS library, now accessible for utilization with various detector geometries and tracking algorithms by a broad community of users.

 In order to perform the optimization studies, $t\bar{t}$ events generated and simulated within the ACTS framework are used. These events correspond to the proton-proton (pp) collisions at a center-of-mass energy $\sqrt{s}=14$~TeV with an additional pile-up (additional pp interactions within the same bunch-crossing) of 200, emulating the High Luminosity LHC conditions. The events are generated using Pythia8~\cite{10.21468/SciPostPhysCodeb.8} and only particles with a transverse momentum p$_{\textrm{T}}$ $>$ 1~GeV have been considered for current studies. 

\section{Parameter Optimization in ACTS}
The automatic parameter optimization techniques featured in this paper have been integrated into the ACTS library. These techniques are now accessible for utilization with various detector geometries and tracking algorithms by a broad community of users.
\subsection{Optimization Framework}
\begin{figure}[!h]
\begin{center}
    \includegraphics[width=0.7\textwidth]{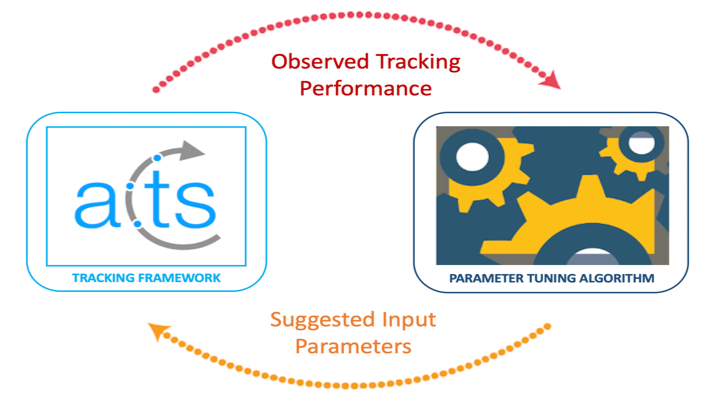}
    %$\includegraphics[width=0.45\textwidth]{Plots/ODD.png}
    \caption{Integration of ACTS tracking framework with parameter tuning framework.}
    \label{fig:ACTS_tuning}
    \end{center}
\end{figure}

Due to the difficulty of finding the derivatives of the tracking algorithms, this work explores derivation-free optimization techniques. The objective is to discover the optimal configuration that yields the best performance. To achieve this, a scoring function is formulated, drawing from various performance metrics of the tracking algorithm, such as efficiency and fake rate. To perform the optimization, the ACTS tracking framework is integrated with parameter tuning framework as presented in figure~\ref{fig:ACTS_tuning}. The tuning framework initiates the process by providing a random input parameter configuration to the tracking framework. Subsequently, the tracking framework assesses its performance based on this configuration and reports the score back to the tuning framework. The tuning framework strives to maximize this score, and consequently, the algorithm's performance. It employs parameter estimation techniques to generate another set of parameters, which are then passed to the tracking algorithm. This iterative process continues for a number of iterations until a satisfactory score and parameter configuration are achieved. 

This study explores two distinct optimization frameworks:
Or\'ion~\cite{xavier_bouthillier_2022_0_2_6}, an asynchronous framework for black-box function optimization, and Optuna~\cite{optuna_2019}, an open source software for automatic hyperparameter search. Both of these algorithms can be easily implemented and integrated within ACTS, making it convenient to use them for the optimization of different algorithms. 
Within the Or\'ion framework, a random search algorithm has been implemented that randomly samples the parameter configurations from the parameter space and provides an optimal configuration for a large enough number of trails. Meanwhile, in the Optuna framework,  The Tree-structured Parzen Estimator (TPE)~\cite{NIPS2011_86e8f7ab} algorithm has been utilized. TPE is a Bayesian optimization method that determines parameter configurations based on a probabilistic model.   

\subsection{Implementing optimization in different ACTS algorithms}
\subsubsection{Material Mapping}
In a real detector, as particles traverse the detector material, they undergo deviations from their initial trajectory due to the interactions with matter. These interactions hold substantial importance in determining the paths of these particles and needs to be taken into account properly when reconstructing the trajectory of the particles. To effectively incorporate the trajectory changes resulting from the material interactions in a virtual detector, a high degree of precision is essential in determining the quantity of material at each point within the detector.

\begin{figure}[!h]
\begin{center}
    \includegraphics[width=0.45\textwidth]{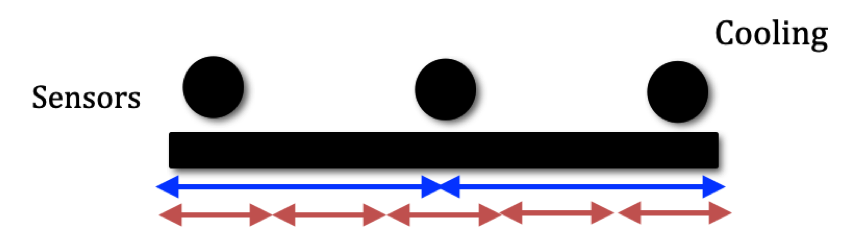}
    \includegraphics[width=0.5\textwidth]{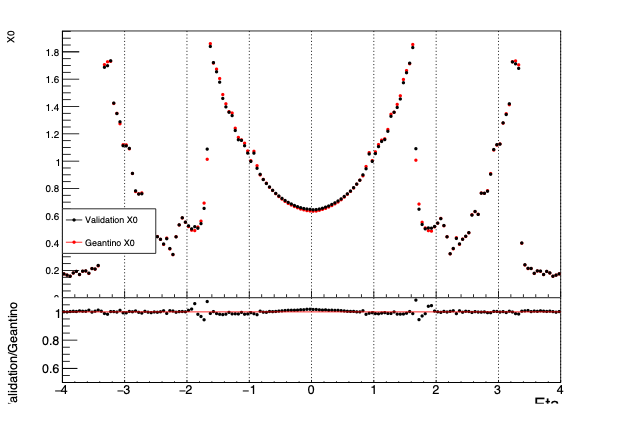}
    \caption{(a) Projection of the material onto bins (arrows) of different sizes. (b) Comparison between the material encountered in a Geant4 simulation vs. the auto-tuned material map in the the open data detector.}
    \label{fig:ACTS_MaterialMapping}
    \end{center}
\end{figure}

Conventional detector simulation frameworks such as Geant4~\cite{AGOSTINELLI2003250} excel in accurately simulating detector materials. However, they demand significant memory resources, and incorporating them into the track reconstruction process would significantly impede the speed of execution. An approximation can be achieved by creating a simplified material map, where all the material is projected onto a predefined set of surfaces within the detector. Each surface is divided into two-dimensional bins, and the material within each bin is averaged. The process is visualized in Figure~\ref{fig:ACTS_MaterialMapping}(a) that provides an illustration of the material mapping on one such surface. As a particle track crosses one of these surfaces, we can identify the specific bin it intersects and compute the energy loss due to interaction by determining the material content stored in that bin. 
The success of this approach depends significantly on the careful selection of specific surfaces and the choice of suitable bin sizes for each surface. Opting for coarse bins can result in the loss of crucial geometric information, leading to biased reconstructions. Conversely, opting for excessively fine bins will inflate memory consumption. Therefore, it is essential to optimize these parameters to create an efficient material map.

We have used automatic optimization algorithm called Or\'ion to optimize the bin size of the surfaces used in the material map. The set of surfaces is provided to the algorithm and the algorithm finds the optimal binning automatically. For this test, we have used a total of 107 surfaces, that leads to a total of 214 parameters. The algorithm uses a random search approach to find the best parameters while trying to minimize a score function (defined in equation(\ref{Score_mat})) which is based on the quality of the map. The $bins$ in the score function refers to the number of bins in a surface while $variance$ refers to the variance of all the material projected onto a given bin. The algorithm tries to minimize the score which in turn minimizes the variance and keeps the number of bins as small as possible. We were able to achieve optimal results on 40 CPU cores after one day of running. The resultant material map showed a good agreement with the map generated by the Geant4 as shown in Figure~\ref{fig:ACTS_MaterialMapping}(b).

\begin{equation}
Score = \frac{1}{bins} \times \sum_{bin}variance_{bin} \times (1+\sqrt{bins})
\label{Score_mat}	
\end{equation}

\subsubsection{Track Seeding}

Reconstructing particle tracks from the detector hits is a complex challenge due to the substantial combinatorial nature of the problem when numerous hits are present in the detector. Therefore, most of the track reconstruction algorithms start with an initial step called the ``Track Seeding''. During this phase, small tracklets, known as track seeds, are formed using only the initial few detector layers (usually three or more) and a helicoidal fit is performed on these tracklets to get a coarse estimate of the corresponding track parameters. Subsequently, these tracklets serve as the input for the complete track-finding process. 

The track seeding algorithm uses a set of user-defined parameters to select hits for seed reconstruction and these parameters can differ significantly depending on the detector geometry and other experimental conditions. Finding the optimal parameter configuration for a given geometry is critical for efficient seed reconstruction, as the failure to reconstruct a seed corresponding to a truth particle can result in the permanent loss of that particle, significantly affecting tracking performance. On the other hand, too many seeds per truth particle can slow down the track reconstruction
process. 

The manual evaluation of these parameters can be a tedious task, so to streamline this process, we have employed auto-tuning techniques, utilizing two distinct optimization algorithms: Optuna and Or\'ion. In this study, we have focused on optimizing eight seeding parameters while assessing the performance using three crucial metrics: the efficiency (the fraction of particles accurately reconstructed as tracks), the fake rate (the fraction of reconstructed tracks not corresponding to any particle) and the duplicate rate (the fraction of reconstructed tracks that duplicate previously reconstructed ones). The first two metrics directly impact the physics performance, while the third metric impact the speed of reconstruction. A scoring function based on these metrics was constructed to guide our optimization efforts: 
	
\begin{equation}
Score = Efficiency - (FakeRate + \frac{DuplicateRate}{K} + \frac{RunTime}{K})
\label{Score_seed}	
\end{equation}

\begin{figure}[!h]
\begin{center}
    \includegraphics[width=0.45\textwidth]{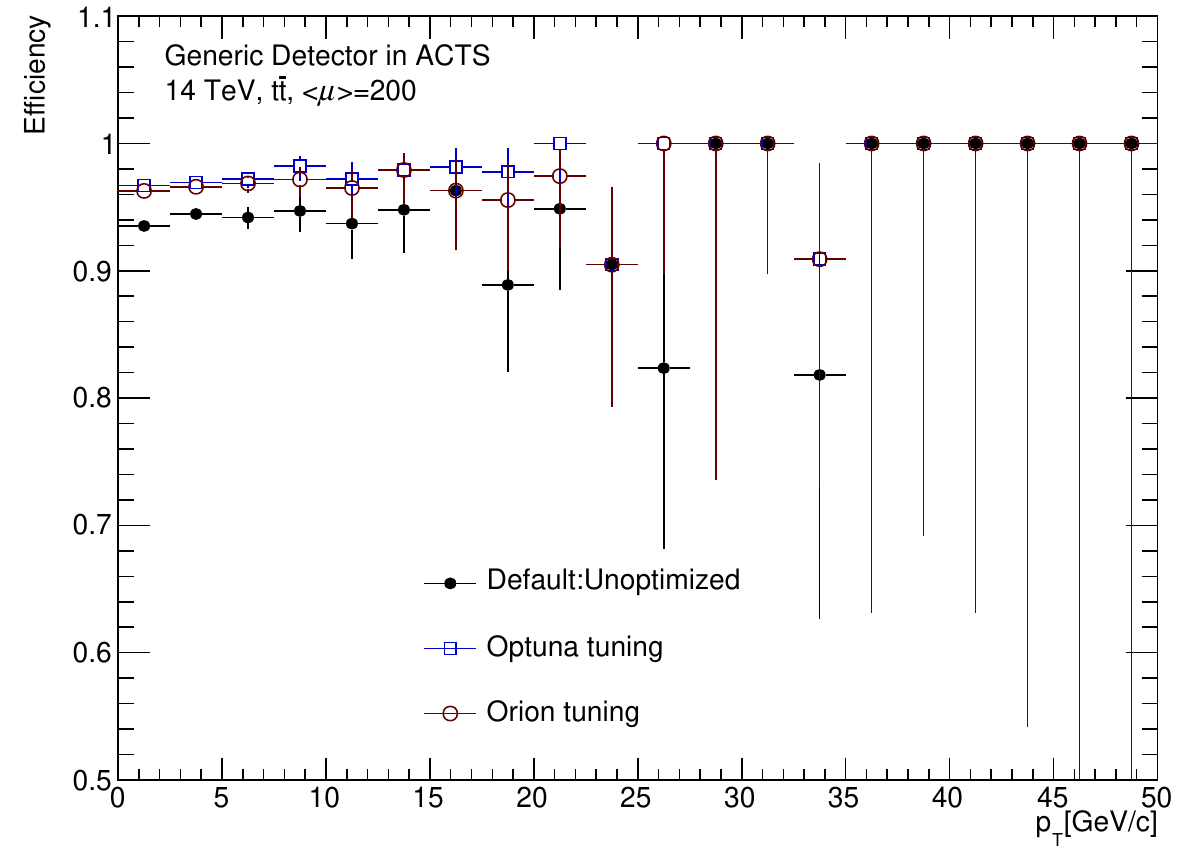}
    \includegraphics[width=0.45\textwidth]{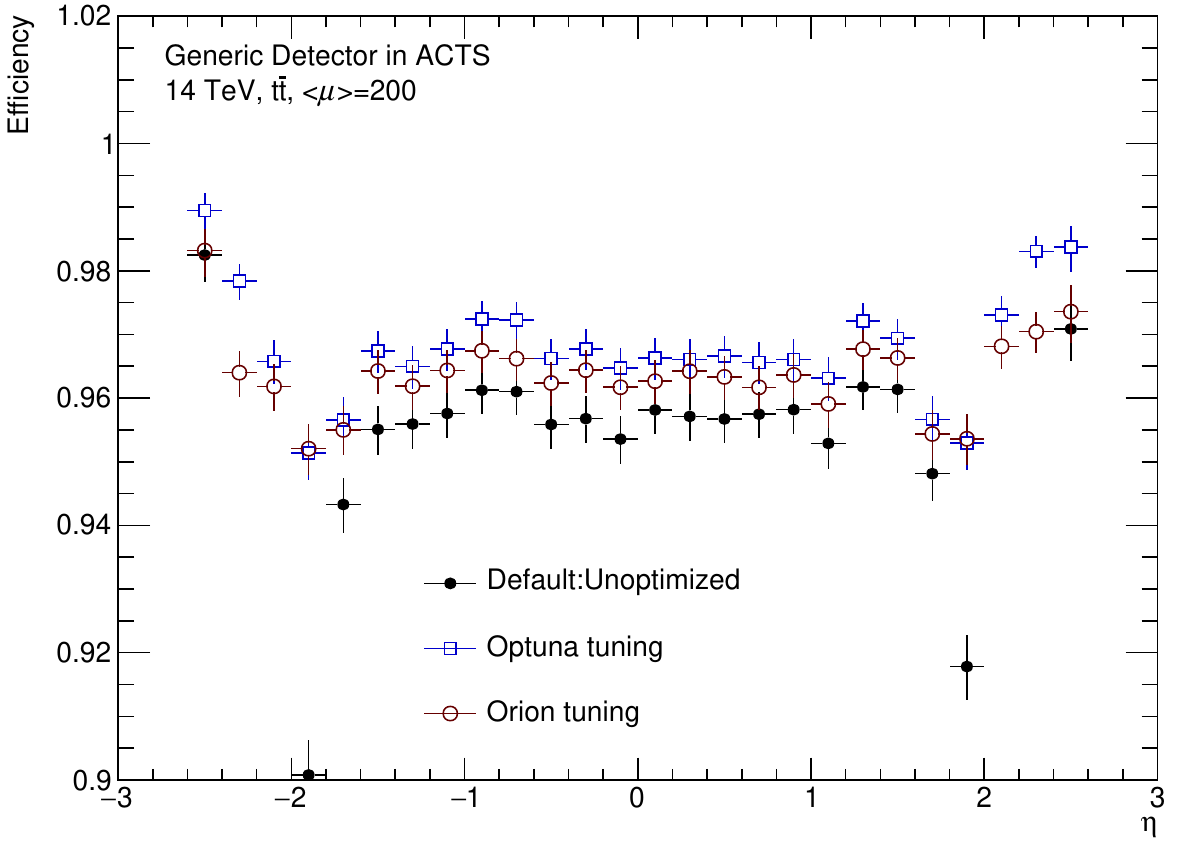} \\
    \includegraphics[width=0.45\textwidth]{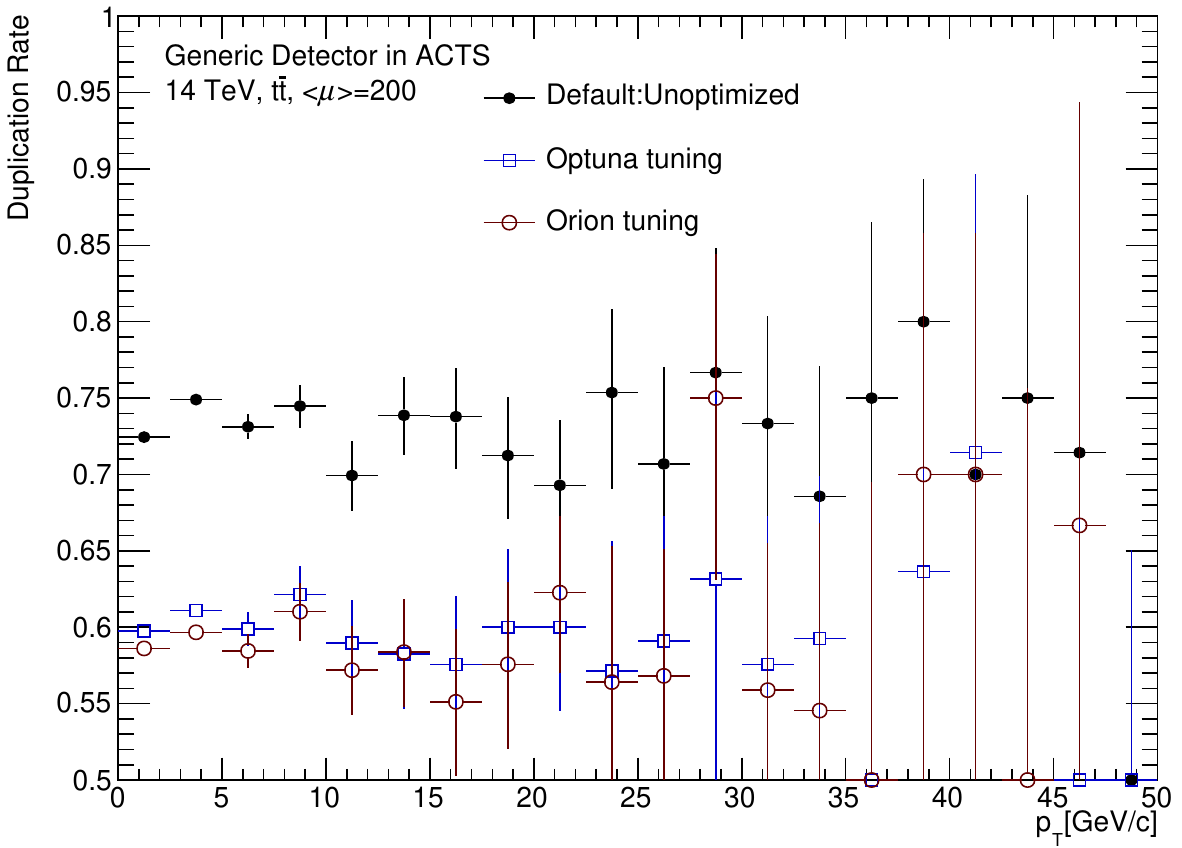}
    \includegraphics[width=0.45\textwidth]{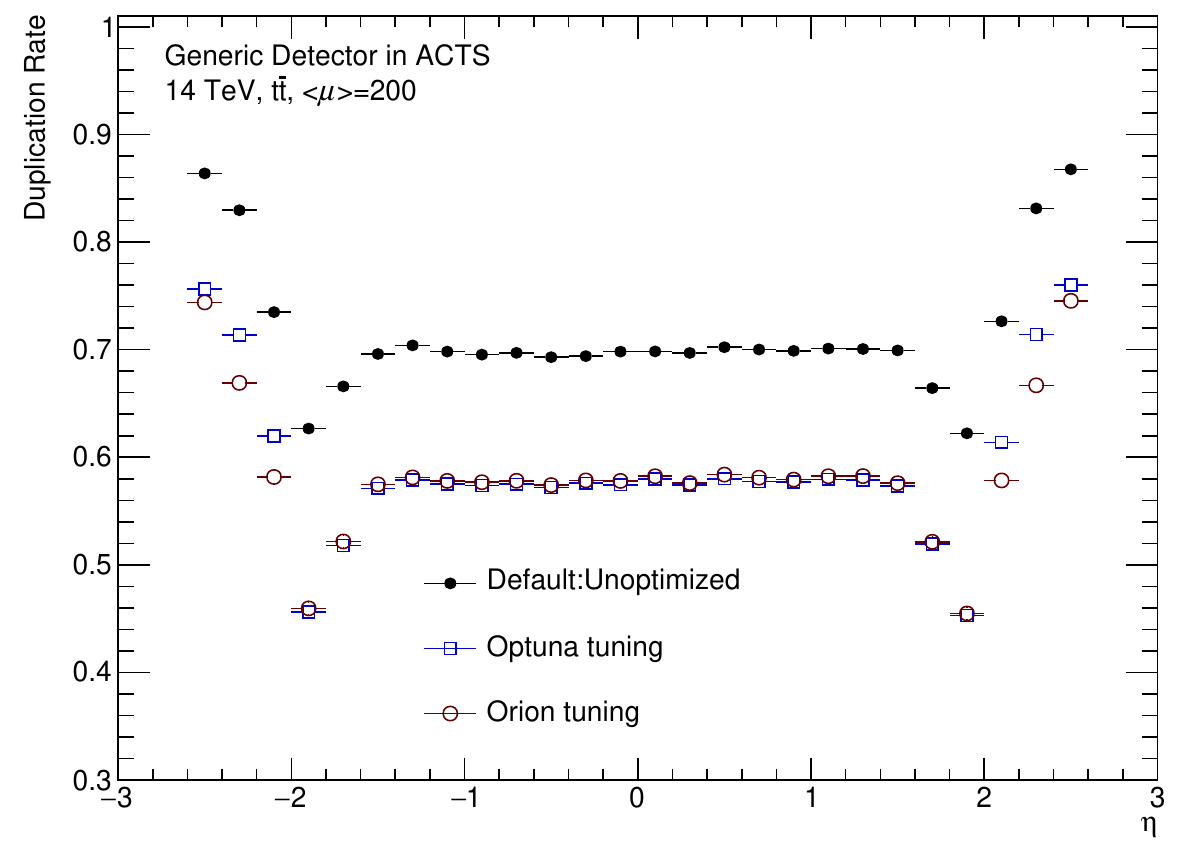} 
    \caption{Comparison of track reconstruction efficiency and duplicate rate before and after parameter optimization. The track reconstruction efficiency as a function of transverse momentum and pseudo-rapidity is shown in top-left and top-right figures while the corresponding distributions for duplicate rate are shown in two bottom figures. The low statistics in high P$_{\textrm{T}}$ regions results in large error-bars.} 
    \label{fig:TrackSeeding}
    \end{center}
\end{figure}

The optimization algorithms try to maximize the score while trying to achieve a better performing parameter configuration. 
Remarkably, both Optuna and  Or\'ion efficiently converged to optimal parameter configurations within one hour of running. Figure~\ref{fig:TrackSeeding} illustrates a comparison of efficiency and duplicate rates before and after optimization. The results demonstrate a notable enhancement compared to an un-optimized configuration in both cases. 

\subsubsection{Vertexing}
In collider experiments, when two beams collide, numerous interactions can occur simultaneously. The interaction characterized by the highest momentum exchange is known as the ``hard interaction'', while all others are categorized as the ``pile-up interactions''. The collision points where these interactions occur are denoted as primary and pile-up vertices, respectively. To differentiate between tracks originating from the primary and pile-up vertices, the tracks are associated with their corresponding vertices using a vertex reconstruction algorithm.  Within the ACTS framework, we employ an algorithm known as the Adaptive Multi-Vertex Finder (AMVF)~\cite{ATL-PHYS-PUB-2019-015}  for this purpose.
The AMVF algorithm concurrently fits multiple tracks to various vertices, assigning different weights to each track-vertex pairing. Subsequently, it optimally fits all the vertices and assigns each track to its respective vertex.

Similar to the track seeding algorithm, the performance of the vertexing algorithm relies on several user-defined parameters that exhibit significant variation depending on the detector geometry and other experimental conditions. To enhance the algorithm's performance, we have employed both Optuna and Or\'ion to obtain optimized parameter configurations.
 For this study, we have chosen five vertexing parameters for optimization. The following score function has been constructed based on a number of performance metrics: 

\begin{equation}
Score = (Eff_{Total}+2Eff_{Cleaned})-(Merged+Split+Fake+Resolution)
\label{Score_vertex}	
\end{equation}
where $Eff_{Total}$ is the fraction of reconstructed truth vertices, $Eff_{Cleaned}$ is the fraction of reconstructed truth vertices such that each reconstructed vertex is associated to only one truth vertex, $Merged$ is the fraction of reconstructed vertices associated with multiple truth vertices, $Split$ is the fraction of reconstructed vertices such that multiple vertices are associated with the same truth vertex, $Fake$ is the fraction of reconstructed vertices not associated to any truth vertex and $Resolution$ is reconstructed vertex resolution in $x, y$ and $z$. 
 
\begin{figure}[!h]
\begin{center}
    \includegraphics[width=0.45\textwidth]{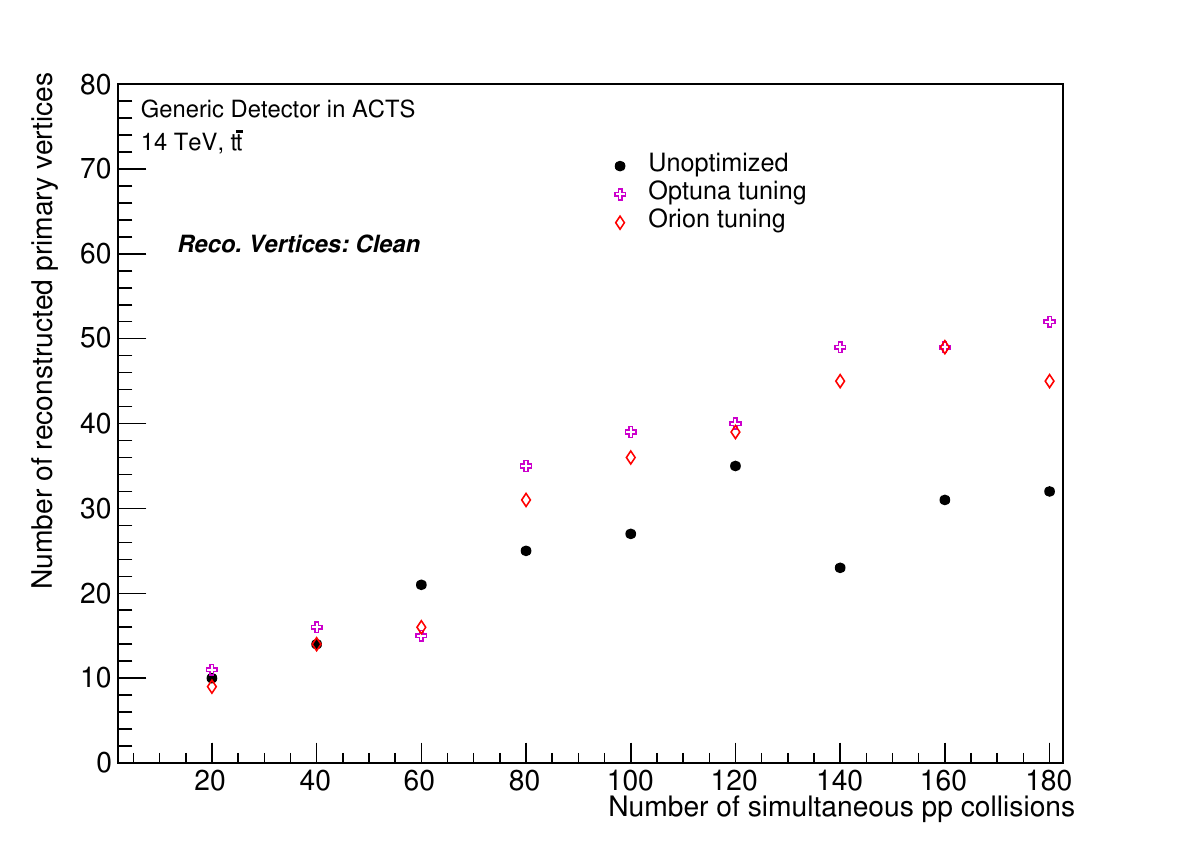}
    \includegraphics[width=0.45\textwidth]{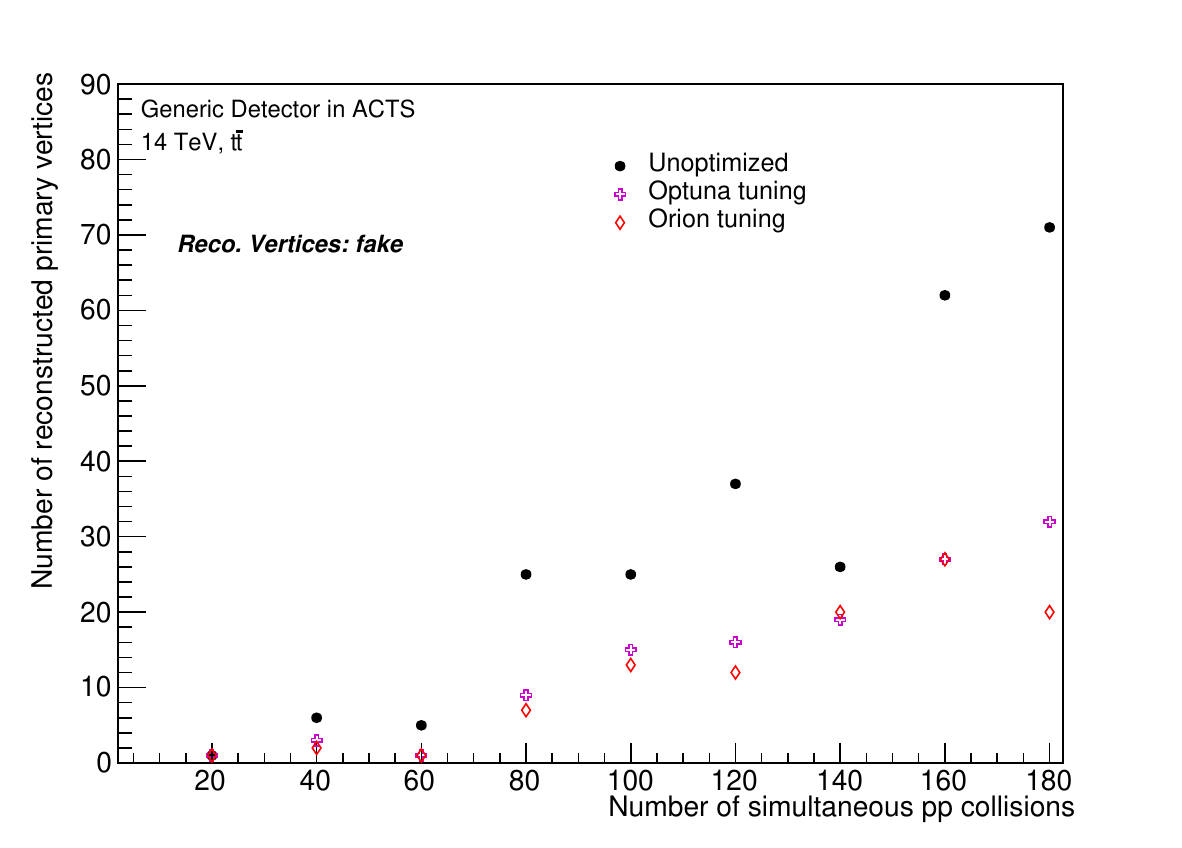}
    \caption{Comparison of the number of reconstructed vertices as a function of pile-up. The left plot shows the number of clean vertices while right plot shows the number of fake vertices.}
    \label{fig:Vertexing}
    \end{center}
\end{figure}

 Our objective for the vertexing algorithm is to maximize the reconstruction of clean vertices while maintaining a high overall efficiency. To attain this objective, our optimization algorithms strive to maximize the score, typically converging to an optimal configuration within approximately four hours of runtime.  The resulting number of clean and fake vertices is shown as a function of the pile-up in Figure~\ref{fig:Vertexing}.
 Notably, substantial improvements in the number of clean vertices are evident, particularly in high pile-up scenarios, when compared to the un-optimized configuration. Likewise, a reduction in the number of fake vertices is also observed.

\section{Conclusion}

Our research demonstrates the effective application of data-driven auto-tuning algorithms within the realm of track reconstruction. We have successfully optimized input parameters for various algorithms, including seed reconstruction, vertex reconstruction, and material mapping. These methods have been seamlessly integrated into the ACTS framework, making them readily available for adoption by any experiment utilizing ACTS for their tracking requirements.

Looking ahead, our ongoing efforts will be directed toward further generalizing this approach. We aim to extend the scope of automatic parameter tuning to encompass a broader range of algorithms within the ACTS framework, simplifying the optimization process for a wider array of tracking tasks.

\section{Acknowledgments}
This project has received funding from the European Unions Horizon 2020 research and innovation programme under grant agreement No 101004761. \\This work was supported by the National Science Foundation under Cooperative Agreements OAC-1836650 and PHY-2323298.
%For tables use syntax in table~\ref{tab-1}.
%\begin{table}
%\centering
%\label{tab-1}       % Give a unique label
% For LaTeX tables you can use
%\begin{tabular}{lll}
%\hline
%first & second & third  \\\hline
%%number & number & number \\\hline
%\end{tabular}
% Or use
%\vspace*{5cm}  % with the correct table height
%\end{table}
%
% BibTeX or Biber users please use (the style is already called in the class, ensure that the "woc.bst" style is in your local directory)
% \bibliography{name or your bibliography database}

\bibliography{template.bib}

\begin{thebibliography}{10}

\bibitem{Ai:2021ghi}
X.~Ai, C.~Allaire, N.~Calace, A.~Czirkos, M.~Elsing, E.~Ene, R.~Farkas, L.G. Gagnon, R.~Garg, P.~Gessinger et~al., Computing and Software for Big Science \textbf{6} (2022)

\bibitem{Amrouche2023}
S.~Amrouche, L.~Basara, P.~Calafiura, D.~Emeliyanov, V.~Estrade, S.~Farrell, C.~Germain, V.V. Gligorov, T.~Golling, S.~Gorbunov et~al., Computing and Software for Big Science \textbf{7} (2023), \texttt{2105.01160}

\bibitem{corentin_allaire_2022_6445359}
C.~Allaire, P.~Gessinger, J.~Hdrinka, M.~Kiehn, F.~Kimpel, J.~Niermann, A.~Salzburger, S.~Sevova, \emph{Opendatadetector} (2022), \urlstyle{tt}\url{https://doi.org/10.5281/zenodo.6445359}

\bibitem{Frank_2014}
M.~Frank, F.~Gaede, C.~Grefe, P.~Mato, Journal of Physics: Conference Series \textbf{513}, 022010 (2014)

\bibitem{10.21468/SciPostPhysCodeb.8}
C.~Bierlich, S.~Chakraborty, N.~Desai, L.~Gellersen, I.~Helenius, P.~Ilten, L.~Lönnblad, S.~Mrenna, S.~Prestel, C.T. Preuss et~al., SciPost Phys. Codebases p.~8 (2022)

\bibitem{xavier_bouthillier_2022_0_2_6}
X.~Bouthillier, C.~Tsirigotis, F.~Corneau-Tremblay, T.~Schweizer, L.~Dong, P.~Delaunay, F.~Normandin, M.~Bronzi, D.~Suhubdy, R.~Askari et~al., \emph{Epistimio/orion: Asynchronous distributed hyperparameter optimization} (2022), \urlstyle{tt}\url{https://doi.org/10.5281/zenodo.3478592}

\bibitem{optuna_2019}
T.~Akiba, S.~Sano, T.~Yanase, T.~Ohta, M.~Koyama, \emph{Optuna: A Next-generation Hyperparameter Optimization Framework}, in \emph{Proceedings of the 25th {ACM} {SIGKDD} International Conference on Knowledge Discovery and Data Mining} (2019)

\bibitem{NIPS2011_86e8f7ab}
J.~Bergstra, R.~Bardenet, Y.~Bengio, B.~K\'{e}gl, \emph{Algorithms for Hyper-Parameter Optimization}, in \emph{Advances in Neural Information Processing Systems}, edited by J.~Shawe-Taylor, R.~Zemel, P.~Bartlett, F.~Pereira, K.~Weinberger (Curran Associates, Inc., 2011), Vol.~24, \urlstyle{tt}\url{https://proceedings.neurips.cc/paper/2011/file/86e8f7ab32cfd12577bc2619bc635690-Paper.pdf}

\bibitem{AGOSTINELLI2003250}
S.~Agostinelli, J.~Allison, K.~Amako, J.~Apostolakis, H.~Araujo, P.~Arce, M.~Asai, D.~Axen, S.~Banerjee, G.~Barrand et~al., Nuclear Instruments and Methods in Physics Research Section A: Accelerators, Spectrometers, Detectors and Associated Equipment \textbf{506}, 250 (2003)

\bibitem{ATL-PHYS-PUB-2019-015}
Tech. rep., CERN, Geneva (2019), all figures including auxiliary figures are available at https://atlas.web.cern.ch/Atlas/GROUPS/PHYSICS/PUBNOTES/ATL-PHYS-PUB-2019-015, \urlstyle{tt}\url{https://cds.cern.ch/record/2670380}

\end{thebibliography}
\end{document}